\documentclass[11pt,a4paper]{article}
\usepackage{t1enc}
\usepackage[latin2]{inputenc}
\usepackage{amsmath}
\usepackage{amsfonts}
\usepackage{amssymb}
\usepackage{amsthm}
\usepackage{graphicx}
\usepackage{color}
\usepackage[english,magyar]{babel}
\title{Finding complex balanced and detailed balanced realizations of chemical reaction networks}
\author{Gábor Szederkényi$^1$, Katalin M. Hangos$^{1,2}$}
\date{}
\newenvironment{gmatrix}[1]{\left[\begin{array}[c]{#1}}{\end{array}\right]}

\newtheorem{theorem}{Theorem}[section]
\newtheorem{definition}{Definition}[section]

\theoremstyle{definition}

\newcommand{\icol}[2]{\ensuremath{\left[ {#1} \right]_{ \cdot, {#2}}}}
\newcommand{\irow}[2]{\ensuremath{\left[ {#1} \right]_{ {#2} , \cdot}}}

\begin{document}
\selectlanguage{english} \maketitle
\begin{center}
$^{1}$
Process Control Research Group\\
Computer and Automation Research Institute,\\
Hungarian Academy of Sciences\\
H-1518, P.O. Box 63, Budapest, Hungary\\
Tel: +36 1 279 6000\\
Fax: +36 1 466 7503\\
e-mail: szeder@sztaki.hu\\
$^{2}$Department of Electrical Engineering and Information Systems, University of Pannonia\\
H-8200, Egyetem u. 10, Veszprém, Hungary
\end{center}

\medskip

\vspace*{2cm}
\begin{center}
\begin{small}
Preprint of the manuscript submitted to \textit{Journal of Mathematical Chemistry}\\ available at \texttt{http://www.springerlink.com}
\end{small}
\end{center}

\newpage
\begin{abstract}
Reversibility, weak reversibility and deficiency, detailed and complex balancing are generally not "encoded" in the kinetic differential equations but they are realization properties 
that may imply local or even global asymptotic stability of the underlying reaction kinetic system when further conditions are also fulfilled. 
In this paper, efficient numerical procedures are given for finding complex balanced or detailed balanced realizations of mass action type chemical reaction networks or kinetic dynamical systems in the framework of linear programming. 
The procedures are illustrated on numerical examples. 
\end{abstract}
\section{Introduction}
Chemical reaction networks form a wide and intensively studied class of nonnegative polynomial systems \cite{Erdi1989}. Beside describing pure chemical reactions, CRNs are often used to model complex biological mechanisms \cite{NicPri:1977}, or models of application fields seemingly far from chemistry such as mechanical or electrical systems \cite{Samardzija1989}. The increasing and extended interest for this field is shown by the fact that numerous surveys and tutorials have been published even in journals where the primary scope is not chemistry \cite{Son:2001, Chellaboina2009, Angeli2009}.

A special subclass of CRNs is the set of reaction networks where the so-called mass action law (MAL) is assumed to be valid. The strongest theoretical results concerning the dynamical properties of CRNs are available for MAL kinetics. Chemical Reaction Network Theory (CRNT) initiated by the results published e.g. in the pioneering papers and lectures \cite{Horn1972, Feinberg1987, Feinberg1988, Feinberg:79} deals with the relations between structures/parameters and dynamical behaviour of CRNs. Ever since, there has been continuous development in this field with significant contributions such as in \cite{Craciun2005, Craciun2006a, Gorban2004, Shinar2010}.

The principle of detailed balance has been applied in thermodynamics for long, which says that the forward and backward rates of a certain molecular process must be equal at the equilibrium \cite{Callen:80}. For chemical reactions, this means that the forward and reverse rates of each reversible elementary reaction step should be equal at the equilibrium state \cite{Lente2010}. In the 1970s, a special set of CRNs called \textit{complex balanced} systems were identified that exhibit "regular" behaviour of the concentration trajectories in the sense that at least local stability of the positive equilibrium points can be proved and certain "exotic" dynamical phenomena such as periodic or chaotic solutions can be ruled out \cite{Horn1972a, Feinberg1972}. 

The widely known fundamental dogma of chemical kinetics says that parametrically and/or structurally different CRNs can give rise to the same mathematical model in ODE form \cite{Schnell2006}.  In other words, seemingly very different CRNs may have exactly the same function in terms of the time evolution of the species concentrations. Such networks will be called \textit{dynamically equivalent} in this paper. 
In \cite{Szederkenyi2009b} the notions of dense and sparse realizations were introduced for dynamically equivalent reaction networks containing the maximal or minimal number of reactions with a fixed set of complexes. Moreover, a computational method in the framework of mixed integer linear programming (MILP) was proposed to compute such realizations. In \cite{Szederkenyi2011}, important properties of dense realizations were described: it was shown that the structure (i.e. the unweighted directed graph) of a dense realization of a given CRN is unique, and the unweighted directed reaction graph of any other realization is the subgraph of the dense realization if the set of complexes is fixed. Furthermore, additional MILP based methods were given for computing reversible CRN realizations, and realizations with the minimal number of complexes. 

Since, similarly to the number of linkage classes, reversibility, weak reversibility and deficiency, detailed and complex balancing are generally also not "encoded" in the kinetic differential equations but they are realization properties, it is of interest whether a kinetic system or a CRN with multiple possible structures has a detailed/complex balanced realization or not. 
The existence of such realization(s) may imply local or even global asymptotic stability of the underlying reaction kinetic system when further conditions are also fulfilled. 

The outline of the paper is the following. In section \ref{sec:basic_notions}, the basic notions in the fields of basic CRN properties, kinetic realizability of polynomial systems, and linear programming (LP) will be summarized. Section \ref{sec:breals_throughLP} contains the main contributions where the computation of CRN realizations with detailed balance and complex balance is casted into LP problems. Section \ref{sec:examples} contains illustrative examples, and finally, the most important conclusions can be found in section \ref{sec:conclusions}. 
\section{Preliminaries}\label{sec:basic_notions}
\subsection{Nonnegative systems}
The concepts and basic results in this subsection are mostly taken from \cite{Chellaboina2009}. A function $f=[f_1~\dots~f_n]^T:[0,\infty)^n\rightarrow\mathbb{R}^n$ is called \textit{essentially nonnegative} if, for all $i=1,\dots,n$, $f_i(x)\ge 0$ for all $x\in[0,\infty)^n$, whenever $x_i=0$. In the linear case, when $f(x)=Ax$, the necessary and sufficient condition for essential nonnegativity is that the off-diagonal entries of $A$ are nonnegative (such a matrix is often called a \textit{Metzler-matrix}). 

Consider an autonomous nonlinear system
\begin{equation}
\dot{x}=f(x),~x(0)=x_0\label{nonlsys}
\end{equation}
where $f:\mathcal{X}\rightarrow \mathbb{R}^n$ is locally Lipschitz, $\mathcal{X}$ is an open subset of $\mathbb{R}^n$  and $x_0\in\mathcal{X}$. Suppose that the nonnegative orthant $[0,\infty)^n=\bar{\mathbb{R}}^n_+\subset \mathcal{X}$. Then the nonnegative orthant is invariant for the dynamics \eqref{nonlsys} if and only if $f$ is essentially nonnegative.
\subsection{Mass action reaction networks}
The overview in this subsection is largely based on \cite{Szederkenyi2009b} and \cite{Szederkenyi2011}.
A CRN obeying the mass action law
 is a closed system where 
 chemical species $\mathbf{X}_i,~i=1,...,n$ take part in $r$ 
 chemical reactions. 
The concentrations of the species denoted by $x_i,~(i=1,...,n)$ form the state vector, i.e. $x_i=[\mathbf{X}_i]$. 
The \emph{elementary reaction steps} have the following form: 
\begin{equation} \label{Eq:irrevreakt}
 \sum_{i=1}^n \alpha_{ij} \mathbf{X}_{i}   \rightarrow
  \sum_{i=1}^n  \beta_{ij} \mathbf{X}_{i},~~j=1,...,r
\end{equation}
where $\alpha_{ij}$ is the so-called \textit{stoichiometric coefficient} 
 of component $\mathbf{X}_{i}$ in the $j$th reaction, and 
 $\beta_{i \ell}$ is the stoichiometric coefficient of the product 
 $\mathbf{X}_{\ell}$. The linear combinations of the species in eq. (\ref{Eq:irrevreakt}), namely $\sum_{i=1}^{n} \alpha_{ij}\mathbf{X}_{i}$ and $\sum_{i=1}^{n}
\beta_{ij}\mathbf{X}_{i}$ for $j=1,\dots,r$ are called the \textit{complexes} and are denoted by $C_1, C_2, \dots, C_m$.
Note that \textit{the stoichiometric coefficients are always nonnegative 
 integers in classical reaction kinetic systems}. 
The \textit{reaction rates} of the 
 individual reactions can be described as 
\begin{equation} \label{Eq:MALrate} 
 \rho_j(x) = k_j \prod_{i=1}^{n} [\mathbf{X}_{i}]^{\alpha_{ij}} = 
  k_j \prod_{i=1}^{n} x_i^{\alpha_{ij}}~~,~~j=1,...,r
\end{equation}
where $k_j>0$ is the \textit{reaction rate coefficient} 
 of the $j$th reaction. The reaction rate and reaction rate coefficient of the reaction $C_i \longrightarrow C_j$ will also be denoted by $\rho_{ij}(x)$ and $k_{ij}$, respectively, whenever it is more convenient.

In our computations, the following form will be used for the description of the dynamics of CRNs obeying the mass action law \cite{Feinberg:79}: 
\begin{equation}
\dot{x}=Y \cdot A_k\cdot \psi(x)\label{eq:Feinberg_desc}
\end{equation}
where $x=[\mathbf{X}] \in\mathbb{R}^n$ is the concentration vector of the species, $Y\in\mathbb{R}^{n\times m}$ stores the stoichiometric composition of the complexes, $A_k\in\mathbb{R}^{m\times m}$ contains information about the structure of the reaction network, and $\psi:\mathbb{R}^n\mapsto\mathbb{R}^m$ is a monomial-type vector mapping given by
\begin{equation}
\psi_j(x)=\prod_{i=1}^n x_i^{\alpha_{ij}},~~~j=1,\dots,m \label{eq:psi_Y}
\end{equation} 
where $\alpha_{ij}=[Y]_{ij}$. To denote the right hand side of eq. \eqref{eq:Feinberg_desc}, the notation $M=Y\cdot A_k$ will also be used.
The structure of $Y$ and $A_k$ is the following. The $i$th column of $Y$ contains the composition of complex $C_i$, i.e. $Y_{ji}$ is the stoichiometric coefficient of $C_i$ corresponding to the specie $\mathbf{X}_j$. $A_k$ is a column conservation matrix (i.e. the sum of the elements in each column is zero), called the \textit{Kirchhoff matrix} of the CRN, defined as
\begin{equation}
[A_k]_{ij}=\left\{
\begin{array}{ccc}
-\sum_{l=1}^m k_{il}, & \text{if} & i=j \\
k_{ji}, & \text{if} & i\ne j
\end{array}
\right.
\end{equation}
It is important to note that the pair $(Y,A_k)$ uniquely characterizes 
 a particular realization of a given CRN. The invariance of the nonnegative orthant for the dynamics of CRNs is shown by several authors (see, e.g. \cite{Erdi1989, Chellaboina2009}).

 To handle the exchange of materials between the environment and the reaction network, the so-called "zero-complex" can be introduced and used which is a special complex where all stoichiometric coefficients are zero i.e., it is represented by a zero vector in the $Y$ matrix \cite{Feinberg:79}. 

For the reaction $C_i\rightarrow C_j$, the corresponding \textit{reaction vector}  is defined as
\begin{equation}
v_k=[Y]_{\cdot, j} - [Y]_{\cdot, i}\label{eq:reacto_vect}
\end{equation}
where $[Y]_{\cdot, i}$ denotes the $i$th column of $Y$. Similarly to reaction rate coefficients, whenever it is more practical, $v_{ij}$ denotes the reaction vector corresponding to the reaction $C_i\rightarrow C_j$.

The \textit{rank} of a reaction network denoted by $s$ is defined as the rank of the vector set $H=\{v_1,v_2\dots,v_r \}$ where $r$ is the number of reactions. The elements of $H$ span the so-called \textit{stoichiometric subspace} denoted by $S$, i.e. $S=\text{span}\{v_1,v_2\dots,v_r \}$. The  positive \textit{stoichiometric compatibility class} containing a concentration $x_0$ is the following set \cite{Feinberg1987}:
\[
(x_0+S)\cap\mathbb{R}^n_+
\] 
where $\mathbb{R}^n_+$ denotes the positive orthant in $\mathbb{R}^n$.

The deficiency $d$ of a reaction network is defined as \cite{Feinberg:79, Feinberg1987}
\begin{equation}
d=m-l-s
\end{equation}
where $m$ is the number of complexes in the network, $l$ is the number of linkage classes (graph components) and $s$ is the rank of the reaction network.

Similarly to \cite{Feinberg:79} and many other authors, the following weighted directed graph (often called \textit{Feinberg-Horn-Jackson} graph) is assigned to the reaction network \eqref{Eq:irrevreakt}.
The directed graph $D=(V_d,E_d)$ of a reaction network consists of a finite nonempty set $V_d$ of vertices and a finite set $E_d$ of ordered pairs of distinct vertices called directed edges. The vertices correspond to the complexes, i.e. $V_d=\{C_1,C_2,\dots C_m\}$, while the directed edges represent the reactions, i.e. $(C_i,C_j)\in E_d$ if complex $C_i$ is transformed to $C_j$ in the reaction network. The reaction rate coefficients $k_j$ for $j=1,\dots,r$ in \eqref{Eq:MALrate} are assigned as positive weights to the corresponding directed edges in the graph. A reaction network is called \textit{reversible}, if whenever it contains the reaction $C_i\longrightarrow C_j$, then the reverse reaction $C_j\longleftarrow C_i$ is also present in the CRN. A reaction network is called \textit{weakly reversible}, if each complex in the reaction graph lies on at least one directed cycle (i.e. if complex $C_j$ is reachable from complex $C_i$ on a directed path in the reaction graph, then $C_i$ is reachable from $C_j$ on a directed path). 
An important point of the well-known \textit{Deficiency Zero Theorem} \cite{Feinberg1987} says that the ODEs of a weakly reversible deficiency zero CRN are globally stable with a known logarithmic Lyapunov function for all positive values of the reaction rate coefficients.

If the corresponding forward and backward reaction pairs form a cycle in the directed graph of a reversible CRN, then this cycle of reversible edge-pairs will be called a \textit{circuit} (to distinguish it from a directed cycle). E.g. the CRN in Fig. \ref{dyneq_db_reals} c) contains a circuit of length 3, while the CRNs in Figs. \ref{dyneq_db_reals} a) and b) do not contain any circuit.

\subsection{Kinetic realizability of nonnegative polynomial systems}
An autonomous polynomial nonlinear system of the form \eqref{nonlsys}
is called \textit{kinetically realizable} or simply \textit{kinetic}, if a mass action reaction mechanism given by eq. \eqref{eq:Feinberg_desc} can be associated to it that exactly realizes its dynamics, i.e. $f(x)=Y\cdot A_k \cdot \psi(x)$ where matrix $Y$ has nonnegative integer elements, $\psi$ contains the monomials according to \eqref{eq:psi_Y}, and $A_k$ is a valid Kirchhoff matrix. In such a case, the pair $(Y,A_k)$ will be called a \textit{realization} of the kinetic system \eqref{nonlsys}. As it is expected from linear algebra, the same polynomial system may have many parametrically and/or structurally different realizations, and this is particularly true for kinetic systems coming from application domains other than chemistry.
The problem of kinetic realizability of polynomial vector fields was first examined and solved in \cite{Hars1981} where the constructive proof contains a realization algorithm that produces the weighted directed graph of a possible associated mass action mechanism. According to \cite{Hars1981}, the necessary and sufficient condition for kinetic realizability is that all coordinates functions of the right hand side of \eqref{nonlsys} must have the form
\begin{equation}
f_i(x) = -x_i g_i(x) + h_i(x),~i=1,\dots,n \label{react_real_form}
\end{equation}
where $g_i$ and $h_i$ are polynomials with nonnegative coefficients. Roughly speaking, condition \eqref{react_real_form} means that kinetic systems cannot contain negative cross-effects. From this, it is easy to see that all nonnegative linear systems are kinetic, since a linear system characterized by a Metzler matrix where only the diagonal elements can be negative is obviously in the form of \eqref{react_real_form}. Moreover, classical Lotka-Volterra (LV) systems that are known to be nonnegative, 
are always kinetic, too. However, there are many essentially nonnegative polynomial systems that are not directly kinetic, since some of the monomials in $f_i$ that do not contain $x_i$ have negative coefficients. 
To circumvent this problem, one possible way is to embed the differential equations into the LV class that can be done algorithmically \cite{HerFai:95}. However, this solution usually results in the significant dimension increase of the state space (i.e. in the increase of species in the corresponding kinetic realization). If the equations and the possible initial conditions guarantee that all state variables remain strictly positive throughout the solutions, then a more advantageous method is a simple state-dependent time-rescaling of the equations that preserves the important qualitative properties of the system and the solutions \cite{Szederkenyi2005, Hangos2011}. The above summary clearly shows that the class of systems that are kinetic or can be transformed to kinetic form is really wide within nonnegative systems.

The very short description of the realization algorithm presented in \cite{Hars1981} is the following. Let us write the polynomial coordinate functions of the right hand side of a  kinetic system \eqref{nonlsys} as
\begin{align}
f_i(x)=\sum_{j=1}^{r_i} m_{ij} \prod_{k=1}^{n}x^{b_{jk}},~~i=1,\dots,n
\end{align}
where $r_i$ is the number of monomial terms in $f_i$. Let us denote the transpose of the $i$th standard basis vector in $\mathbb{R}^n$ as $e_i$ and let $B_j=[b_{j1}~\dots~b_{jn}]$. Then, the necessary steps (transcribed into an easily implementable form) for constructing a realization are given by:
\begin{center}
\framebox{
\parbox[c]{12cm}{
\textbf{Algorithm 1} from \cite{Hars1981}\\
For each $i=1,\dots,n$ and for each $j=1,\dots,r_i$ do:
\begin{center}
\parbox[r]{11cm}{
\begin{enumerate}
\item $C_j$ = $B_j + \text{sign}(m_{ij})\cdot e_i$
\item Add the following reaction to the graph of the realization
\[
\sum_{k=1}^n b_{jk}\mathbf{X}_k \longrightarrow \sum_{k=1}^n c_{jk} \mathbf{X}_k
\]
with reaction rate coefficient $|m_{ij}|$, where $C_j=[c_{j1}~\dots~c_{jn}]$.
\end{enumerate}}
\end{center}
}}
\end{center}
As we can see later in section \ref{sec:examples}, \textbf{Algorithm 1} often produces a non-minimal reaction structure with more reactions and/or complexes than the minimal number needed for the kinetic representation of the studied polynomial system. 
\subsection{Complex balance and detailed balance}
\begin{definition}
A CRN realization $(Y,A_k)$ is called complex balanced at $x^*\in\mathbb{R}^n_+$, if 
\begin{align}
 A_k\psi(x^*)=0.
\end{align}
\end{definition}
\begin{definition}
A reversible CRN realization $(Y,A_k)$ is called detailed balanced at $x^*\in\mathbb{R}^n_+$, if 
\begin{align}
\rho_{ij}(x^*)=\rho_{ji}(x^*),~~\forall i,j~\text{such that}~C_i\leftrightarrows C_j~\text{exists}\label{eq:detbal_condition}
\end{align}
i.e. the forward and reverse reaction rates for each reversible reaction are equal at $x^*$.  
\end{definition}
Clearly, if a CRN is complex balanced at $x^*$, then $x^*$ is an equilibrium point, i.e. $YA_k\psi(x^*)=0$. Furthermore, it is easy to check that detailed balancing at $x^*$ implies complex balancing at $x^*$ for reversible networks, but the converse is generally not true.
\begin{definition}
A reversible CRN realization $(Y,A_k)$ is called complex balanced (detailed balanced), if it is complex balanced (detailed balanced) at each positive steady state.
\end{definition}
It is important to remark, that both the complex balance and detailed balance properties depend on the structure of the reaction digraph and on the numerical values of the reaction rate coefficients \cite{Craciun2009}. In \cite{Feinberg1989}, necessary and sufficient conditions were given for detailed balancing (see also \cite{Nagy2009} for a clear explanation of conditions) using the circuit and spanning forest conditions below. The \textit{circuit conditions} require that the product of the reaction rate coefficients should be equal taken in either directions along a circuit. The \textit{spanning forest conditions} say that 
\begin{align}
\prod (k_{ij})^{\gamma_{ij}} = \prod (k_{ji})^{\gamma_{ij}}
\end{align}
where the products are taken for each reaction step in the spanning forest of the reaction digraph and
$\gamma$ denotes the solutions of the equation 
\begin{align}
\sum_{i,j}\gamma_{ij} v_{ij}=0,~~\text{for all $i,j$ such that $C_i\longrightarrow C_j$ exists},
\end{align}
and $v_{ij}$ is the reaction vector of the reaction $C_i\longrightarrow C_j$. 

It is worth summarizing the following important properties and relations of detailed and complex balancing collected from \cite{Feinberg1972, Horn1972a, Feinberg1989, Gunawardena2003, Craciun2009, Dickenstein2010}.
\begin{description}
\item[P1] If there is no circuit in the reaction graph, then the spanning forest condition is alone necessary and sufficient for detailed balancing.
\item[P2] If the deficiency of the CRN is zero, then the circuit condition is necessary and sufficient for detailed balancing. 
\item[P3] If the deficiency of the CRN is zero, and there is no circuit in the reaction digraph, then the reaction is detailed balanced. 
\item[P4] If the circuit conditions are satisfied in a reversible CRN containing at least one circuit, then detailed balancing and complex balancing are equivalent.
\item[P5] If a CRN is complex balanced (detailed balanced) at any positive $x^*$ then it is complex balanced (detailed balanced) at all other positive equilibrium points.
\item[P6] If a CRN is complex balanced then it is weakly reversible.
\item[P7] A CRN is complex balanced for any positive values of reaction rate coefficients if and only if a) it is weakly reversible, and b) the deficiency of the system is zero.
\item[P8] If a CRN is complex balanced then there is precisely one equilibrium point in each stoichiometric compatibility class. Furthermore, each equilibrium point is at least locally stable with a known strict Lyapunov function. 
\end{description}
Additionally, we have to mention the so-called \textit{Global Attractor Conjecture} for which no counterexample has been found, but the construction of a complete proof seems to be technically challenging \cite{Craciun2009}. The conjecture says that for any complex balanced CRN and any initial condition $x(0)$, the equilibrium point $x^*$ is a global attractor in the corresponding positive stoichiometric compatibility class. The conjecture further supports the need for a method to find complex balanced realizations of kinetic systems, if stability properties of such systems are to be studied. The circuit and spanning tree conditions for detailed balancing are hard to insert into an optimization framework in their original form, but another simple algebraic condition will allow the use of linear programming for this purpose. Moreover, property \textbf{P5} will enable us to use an arbitrary positive steady state to find a detailed balanced or complex balanced realization if it exists.
\subsection{Linear programming}
Linear programming (LP) is an optimization technique, where a linear objective function is minimized/maximized subject to linear equality and inequality constraints \cite{Dantzig1997, Dantzig2003}. LP has been widely used in chemistry and chemical engineering in the fields of system analysis \cite{Belov2010, Kauchali2002}, simulation \cite{Gopal1997} and design \cite{Klein1992}. The standard form of linear programming problems that will be used in this paper is the following
\begin{align}
& \text{minimize~~} c^Ty \label{eq:LP_1}\\
& \text{subject to:~~} \nonumber\\
& ~~ A y = b \label{eq:LP_2}\\
& ~~~  y \ge 0 \label{eq:LP_3}
\end{align}
where $y\in\mathbb{R}^n$ is the vector of decision variables, $c\in\mathbb{R}^n$ $A\in\mathbb{R}^{p\times n}$, $b\in\mathbb{R}^{p}$ are known vectors and matrices, and '$\ge$' in \eqref{eq:LP_3} means elementwise nonstrict inequality.

The feasibility of the simple LP problem \eqref{eq:LP_1}-\eqref{eq:LP_3} can be checked using the following necessary and sufficient condition \cite{Dantzig1997}. 
\begin{theorem}\label{th:lpfeasibility}
Consider the auxiliary LP problem
\begin{align}
& \text{minimize~~}\sum_{i=1}^p z_i \label{eq:auxlp_1}\\
& \text{subject to:~~} \nonumber\\
& ~~ A y + z = b \label{eq:auxlp_2}\\
& ~~~  y \ge 0, z\ge 0 \label{eq:auxlp_3}
\end{align}
where $z\in\mathbb{R}^p$ is a vector of auxiliary variables. Let us assume (without the loss of generality) that $b\ge 0$. There exists a feasible solution for the LP problem \eqref{eq:LP_1}-\eqref{eq:LP_3} if and only if the auxiliary LP problem \eqref{eq:auxlp_1}-\eqref{eq:auxlp_3} has optimal value 0 with $z_i=0$ for $i=1,\dots,p$.
\end{theorem}
It is easy to see that $y=0$, $z=b$ is always a feasible solution for \eqref{eq:auxlp_1}-\eqref{eq:auxlp_3}. The above theorem will be useful for establishing that no detailed or complex balanced realization exists in certain cases.

\section{Finding balanced realizations using linear programming}\label{sec:breals_throughLP}
In this section, we assume that an initial CRN realization $(Y^{(1)}, A_k^{(1)})$ or a kinetic polynomial system is given together with an arbitrary positive steady state $x^*$ that has been determined analytically or simply through simulations. If we start from a kinetic polynomial system, we use \textbf{Algorithm 1} for generating an initial realization $(Y^{(1)}, A_k^{(1)})$. In any case, $M^{(1)}=Y^{(1)}\cdot A_k^{(1)}$. Let us denote $\psi(x^*)$ simply by $\psi^*$.
\subsection{Constraints representing mass action dynamics}
Similarly to \cite{Szederkenyi2009b}, let us introduce the following notations. The Kirchhoff-matrix of the searched realization is written as
\begin{align}
A_k^{(s)} = \begin{gmatrix}{rrrr}
-a_{11} & a_{12} & \dots & a_{1m} \\
a_{21} & -a_{22} & \dots & a_{2m} \\
\vdots & \vdots & \vdots & \vdots \\
a_{m1} & a_{m2} & \dots & -a_{mm}
\end{gmatrix}\label{eq:searched_ak}
\end{align}
Then, the optimization variable, $y$ is defined columnwise as 
\begin{align}
y=\begin{gmatrix}{lllllllll}
a_{11} & a_{21} & a_{31} & \dots & a_{m1} & a_{12} & \dots & a_{(m-1)m} & a_{mm}
\end{gmatrix}^T\in\mathbb{R}^{m^2}\label{eq:ydef}
\end{align}
Taking into consideration the diagonal elements in \eqref{eq:searched_ak} with a negative sign enables us to constrain all elements of $y$ to be nonnegative. Let $\irow{W}{i}$ and $\icol{W}{j}$ denote the $i$th row and $j$th column of an arbitrary matrix $W$, respectively. Let us define the following matrices (with known entries) as
\begin{align}
\bar{Y}^i = \begin{gmatrix}{ccccccc}
\icol{Y^{(1)}}{1} & \dots & \icol{Y^{(1)}}{i-1} & -\icol{Y^{(1)}}{i} & \icol{Y^{(1)}}{i+1} & \dots & \icol{Y^{(1)}}{m} \\
1 & \dots & 1 & -1 & 1 & \dots & 1 
\end{gmatrix}\in\mathbb{R}^{(n+1)\times m}, 
\end{align}
\begin{align}
\bar{M} = \begin{gmatrix}{c}
M^{(1)} \\ 0~\dots~0
\end{gmatrix}\in\mathbb{R}^{(n+1)\times m}, 
\end{align}
Then the equality constraints encoding mass action dynamics can be written as
\begin{align} \label{Eq:condMAL}
A_{e1}\cdot y = B_{e1}
\end{align}
where
\begin{align}
A_{e1}=\begin{gmatrix}{ccccc}
\bar{Y}^1 & 0 & 0 & \dots & 0 \\
0 & \bar{Y}^2 & 0 & \dots & 0 \\
\vdots & \vdots & \vdots & & \vdots \\
0 & 0 & 0 & \dots & \bar{Y}^m
\end{gmatrix}\in\mathbb{R}^{m(n+1)\times m^2}
\end{align}
(with zeros representing zero blocks of size $(n+1)\times m$), and
\begin{align}
B_{e1}=\begin{gmatrix}{c}
\icol{\bar{M}}{1} \\
\icol{\bar{M}}{2} \\
\vdots\\
\icol{\bar{M}}{m} 
\end{gmatrix}
\end{align}
The inequalities expressing sign constraints for the reaction rate coefficients are simply
\begin{align} \label{Eq:condsign}
 y \ge 0.
\end{align}

\paragraph*{The choice of the objective function}
In principle, the objective function (\ref{eq:LP_1}) can be any linear function of $y$ 
that is determined by the choice of the vector $c$. 
In section \ref{sec:examples}, we will minimize or maximize the sum of reaction rate coefficients. The minimization of the sum corresponds to the 
choice of $c=\frac{1}{2}[1~\dots~1]^{T}$ in the standard LP-problem (\ref{eq:LP_1})-(\ref{eq:LP_3}), 
while the maximization is obtained by selecting $c=-\frac{1}{2}[1~\dots~1]^{T}$. 
Note that the minimum of the sum of reaction rate coefficients is bounded if the LP constraints are feasible, but the maximum is not necessarily, and this must be kept in mind.
\subsection{Additional constraints for complex balancing}
Using the definition of a complex balanced steady state, the equality constraints for complex balancing are
\begin{align}
-a_{11}\psi_1^* + a_{12} \psi_2^* + \dots + a_{1m}\psi_m^* = 0\\
a_{21}\psi_1^* -a_{22} \psi_2^* + \dots + a_{2m}\psi_m^* = 0 \\
\vdots \nonumber\\
a_{m1}\psi_1^* + \dots + a_{m(m-1)}\psi_{m-1}^*- a_{mm}\psi_m^* = 0
\end{align}
Clearly, these constraints can be written as
\begin{align} \label{Eq:condcomp}
A_{e2}\cdot y = B_{e2},
\end{align}
where
\begin{align}
A_{e2} = \begin{gmatrix}{ccccccc}
-\psi_1^* & 0^{m-1} & \psi_2^* & 0^{m-1} & \dots & \psi_m^* & 0^{m-1} \\
0 & \psi_1^* & 0^{m-1} & -\psi_2^* & \dots & \psi_m^* & 0^{m-2} \\
\vdots \\
0^{m-1} & \psi_1^* & 0^{m-1} & \psi_2^* & \dots & 0^{m-1} & -\psi_m^*
\end{gmatrix},~~B_{e2}=0\in\mathbb{R}^m,
\end{align}
and $0^k$ denotes a $1 \times k$ row vector of zeros.
\subsection{Further constraints for detailed balancing}\label{sec:detbal}
The most suitable form of detailed balancing constraints is taken from \cite{Yang2006}. According to this, a given steady state $x^*$ is detailed balancing if and only if
\begin{align}
G\cdot A_k^T = A_k\cdot G \label{eq:detbalcond}
\end{align}
where $G=\text{diag}(\psi^*)$. It is easy to see that \eqref{eq:detbalcond}, if satisfied, implies reversibility of the obtained reaction network. Eq. \eqref{eq:detbalcond} encodes a maximum of $\frac{m(m-1)}{2}$ independent equations that can be written into the linear programming problem as
\begin{align} \label{Eq:conddet}
\psi_i^* y_{(i-1)m+j}-\psi_j^* y_{(j-1)m+i}=0,~~\forall i>j,
\end{align}
since $\left[A_k^{(s)}\right]_{ij}=y_{(j-1)m+i}$ according to eq. \eqref{eq:ydef}.
\section{Examples}\label{sec:examples}
\subsection{Multiple detailed balanced realizations of a simple irreversible network}
\begin{figure}[!htbp]%
\centering
\framebox{\includegraphics[width=3cm]{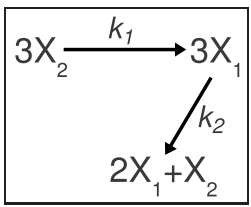}}
\caption{Simple deficiency 1 reaction network}
\label{react_simp1}
\end{figure}
Consider the simple reaction network shown in Fig. \ref{react_simp1} with parameters $k_1=1$, $k_2=1.5$. This CRN was also used as an example in \cite{Szederkenyi2011} in a different context. The matrices characterizing complex composition and graph structure are
\begin{align}
Y=\begin{gmatrix}{ccc}
0 & 3 & 2 \\
3 & 0 & 1
\end{gmatrix},~~A_k=
\begin{gmatrix}{ccc}
-1 & 0 & 0 \\
1 & -1.5 & 0 \\
0 & 1.5 & 0
\end{gmatrix}
\end{align}
Thus,
\begin{align}
M=Y\cdot A_k =
\begin{gmatrix}{ccc}
   3 &  -1.5 &   0 \\
   -3 &    1.5 &    0
\end{gmatrix}.
\end{align} 
The corresponding ODE model is
\begin{align}
\dot{x}_1 &= 3x_2^3 - 1.5 x_1^3 \label{simp_ode1}\\
\dot{x}_2 &= -3x_2^3 + 1.5 x_1^3 \label{simp_ode2}
\end{align}
It's easy to compute that e.g. $x^*=[1.6725~~1.3275]^T$ is a steady state for the system \eqref{simp_ode1}-\eqref{simp_ode2}. The corresponding values of the monomials are $\psi^*=\psi(x^*)=[2.3393 ~~ 4.6786 ~~ 3.7134]^T$. Clearly, the realization shown in Fig. \ref{react_simp1} cannot be complex balanced or detailed balanced, since it is not weakly reversible. 

In the first step, let us define the objective function as
\begin{align}
h(y)=\frac{1}{2}\sum_{i=1}^9 y_i\label{ex2_objfun}
\end{align}
which is the $L1$-norm of the reaction rate coefficient vector (note that the diagonal elements of $A_k$ with a negative sign contain the sum of the reaction rate coefficients in the corresponding column). Minimizing the objective function $h$ in \eqref{ex2_objfun} 
subject to the conditions (\ref{Eq:condMAL}), (\ref{Eq:condsign}) 
 and (\ref{Eq:conddet}) gives the detailed balance solution
\begin{align}
y_{opt}=\begin{gmatrix}{ccccccccc}
1 & 1 & 0 & 0.5 & 0.5 & 0 & 0 & 0 & 0
\end{gmatrix}^T,
\end{align}
that corresponds to the following realization
\begin{align}
Y'=Y,~~A_k'=
\begin{gmatrix}{ccc}
 -1 &  0.5 &    0 \\
    1 &   -0.5 &    0\\
    0  &  0 &   0
\end{gmatrix}
\end{align}
The detailed balance condition at the steady state can be checked as
\begin{align}
1\cdot (x_2^*)^3 = 0.5\cdot (x_1^*)^3 = 2.3393.
\end{align}
which is a reversible, deficiency 0 realization. Therefore, by applying the \textit{Deficiency zero theorem}, the dynamics of the initial irreversible network is also globally stable with a known entropy-like Lyapunov function. 

\begin{figure}[!htbp]%
\centering
\framebox{\includegraphics[width=12cm]{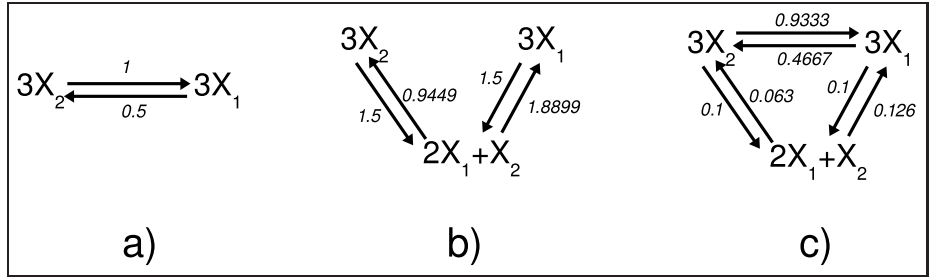}}
\caption{Dynamically equivalent detailed balanced realizations of the CRN shown in Fig. \ref{react_simp1}. a) deficiency zero realization; b), c) deficiency one realizations}
\label{dyneq_db_reals}
\end{figure}

In the second step, let us maximize $h$ in \eqref{ex2_objfun}. The obtained realization is now given by
\begin{align}
Y''=Y,~~A_k''=
\begin{gmatrix}{ccc}
  -1.5 &    0 &    0.9449 \\
    0 &   -1.5 &    1.8899 \\
    1.5 &    1.5 &   -2.8348
\end{gmatrix}
\end{align}
The deficiency of the network in this case is 1. The detailed balance property is fulfilled in this case as
\begin{align}
1.5\cdot (x_2^*)^3 = 0.9449\cdot (x_1^*)^2 x_2^* = 3.5088,\\
1.5\cdot (x_1^*)^3 = 1.8899 \cdot (x_1^*)^2 x_2^* = 7.0179.
\end{align}
The reaction digraphs of the above computed detailed balanced realizations can be seen in Fig. \ref{dyneq_db_reals} a) and b), respectively. 

The motivation for trying to minimize and maximize the $L1$ norm of the reaction rate coefficients to get potentially different detailed balanced realizations came from \cite{Donoho2005} and \cite{Donoho2006}. We remark that the exact conditions under which the sparsest solution of an underdetermined set of linear equations is the minimal $L1$ norm solution are not fulfilled in our case, but the two obtained solutions are structurally different, and our purpose here was only to illustrate that different detailed balanced realizations of the same (very simple) kinetic system may exist.

It's worth mentioning that the detailed balancing constraints described in section \ref{sec:detbal} can easily be combined with the computation of dense and sparse realizations (see \cite{Szederkenyi2009b, Szederkenyi2011}). However, the problem in this case becomes NP-hard. A dense detailed balanced realization can be seen in Fig. \ref{dyneq_db_reals} c) that is determined by the following matrices:
\begin{align}
Y'''=Y,~~A_k'''=
\begin{gmatrix}{ccc}
   -1.0333  &  0.4667  &  0.0630\\
    0.9333 &  -0.5667  &  0.1260\\
    0.1000  &  0.1000  & -0.1890
\end{gmatrix}
\end{align}
The detailed balancing condition is met in this case, too, since
\begin{align}
0.9333 \cdot (x_2^*)^3 = 0.4667 \cdot (x_1^*)^3 = 2.1833 \\
0.1 \cdot (x_2^*)^3 = 0.063 \cdot (x_1^*)^2 x_2^* = 0.2339 \\
0.1 \cdot (x_1^*)^3 = 0.126 \cdot (x_1^*)^2 x_2^* = 0.4679
\end{align}
The average running time of the LP based methods (using the \texttt{linprog} command) was 0.06 second, while the MILP based algorithm (using the freely available GLPK solver) found the dense realization in 1.04 second in the MATLAB computation environment on a notebook computer with an 1.66 GHz Intel Atom N280 Processor.

\subsection{Finding complex balanced realization of a kinetic polynomial system}
\begin{figure}[!htbp]%
\centering
\framebox{\includegraphics[width=5.5cm]{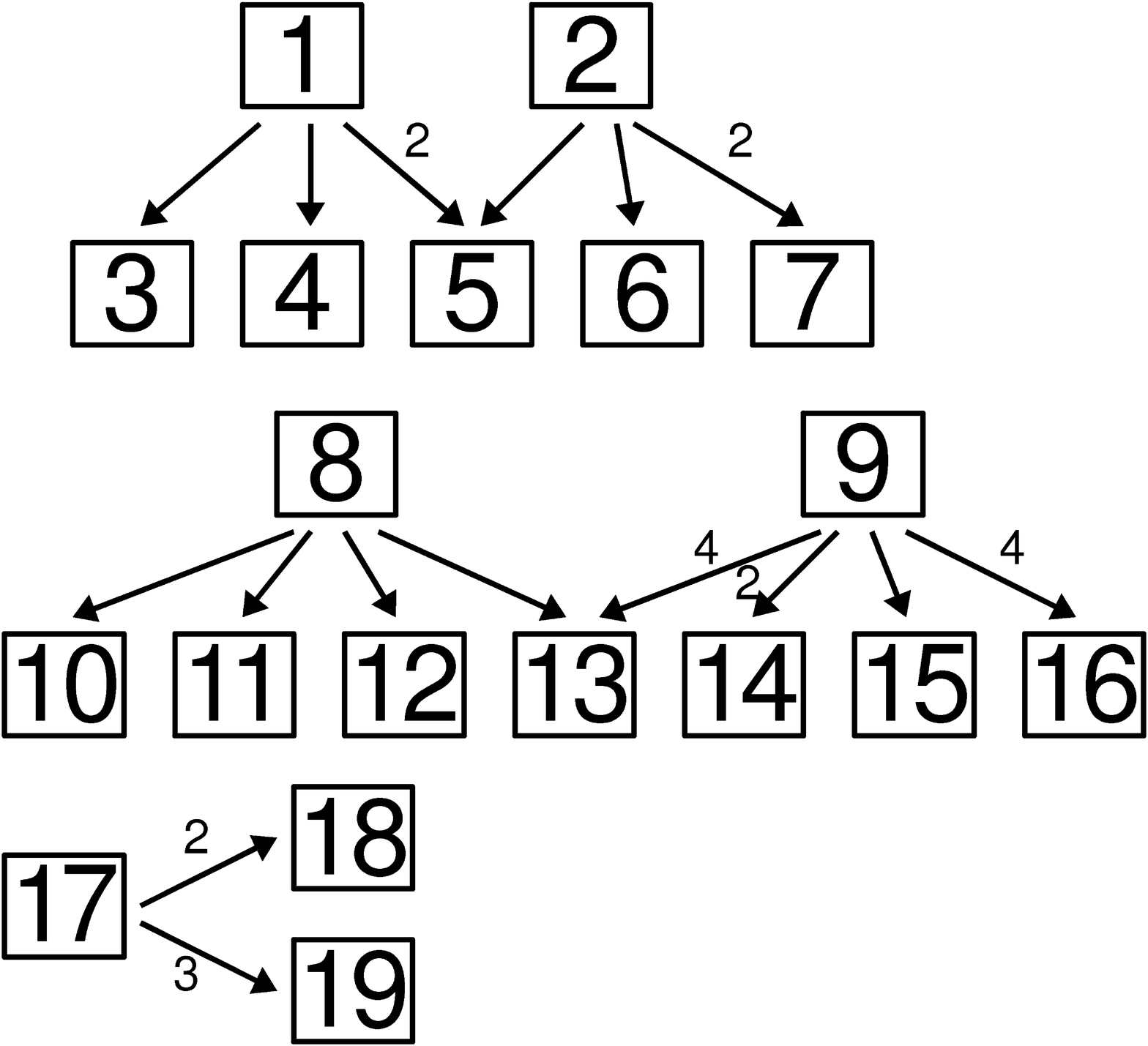}}
\caption{Reaction network realizing eq. \eqref{example2} obtained using \textbf{Algorithm 1}. Only those reaction rate coefficients are indicated that are different from 1.}
\label{react_02}
\end{figure}

\begin{figure}[!htbp]%
\centering
\framebox{\includegraphics[width=4.5cm]{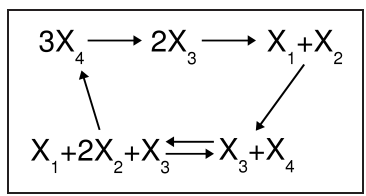}}
\caption{Complex balanced realization of the CRN shown in Fig. \ref{react_02}. All reaction rate coefficients are 1.}
\label{react_02_db}
\end{figure}

The following polynomial system is given:
\begin{small}
\begin{align}
\dot{x}_1 & = x_3^2 - x_1 x_2 + x_3 x_4 - 2 x_1 x_2^2 x_3 \nonumber \\
\dot{x}_2 & = x_3^2 - x_1 x_2 + 2 x_3 x_4 - 4 x_1 x_2^2 x_3 \nonumber \\
\dot{x}_3 & = -2 x_3^2 + x_1 x_2 - x_1 x_2^2 x_3 + 2 x_4^3 \label{example2}\\
\dot{x}_4 & = x_1 x_2 - x_3 x_4 + 4 x_1 x_2^2 x_3 - 3 x_4^3 \nonumber
\end{align}
\end{small}
It can be seen that \eqref{example2} is essentially nonnegative and kinetic. After running \textbf{Algorithm 1}, we obtain an initial network with 19 complexes and 16 reactions that is visible in Fig. \ref{react_02}, where the numbering of complexes (in rectangles) is the following:
\begin{small}
\begin{align} \label{kinsys_complexes}
& 1: 2\mathbf{X}_3,~
2:\mathbf{X}_3+\mathbf{X}_4,~
3:\mathbf{X}_1+2\mathbf{X}_3,~
4:\mathbf{X}_2 + 2\mathbf{X}_3,\nonumber\\
& 5:\mathbf{X}_3,~
6:\mathbf{X}_1 + \mathbf{X}_3 + \mathbf{X}_4,~
7:\mathbf{X}_2 + \mathbf{X}_3 + \mathbf{X}_4,~\nonumber\\
& 8:\mathbf{X}_1+\mathbf{X}_2,~
9:\mathbf{X}_1 + 2\mathbf{X}_2 + \mathbf{X}_3,~
10:\mathbf{X}_1,~
11:\mathbf{X}_2,~\nonumber\\
& 12:\mathbf{X}_1 + \mathbf{X}_2 + \mathbf{X}_4,~
13:\mathbf{X}_1 + \mathbf{X}_2 + \mathbf{X}_3,~
14:2\mathbf{X}_2+\mathbf{X}_3,~\nonumber\\
& 15:\mathbf{X}_1+2\mathbf{X}_2,~
16: \mathbf{X}_1 + 2\mathbf{X}_2 + \mathbf{X}_3 + \mathbf{X}_4,~\nonumber\\
& 17: 3\mathbf{X}_4,~18: \mathbf{X}_3+3 \mathbf{X}_4,~ 19: 2 \mathbf{X}_4 
\end{align}
\end{small}
The objective function that was minimized was also the sum of reaction rate coefficients, i.e.
\begin{align}
h(y)=\frac{1}{2}\sum_{i=1}^{361} y_i 
\end{align}
The constraints were given by eqs. \eqref{Eq:condMAL}, \eqref{Eq:condsign} and \eqref{Eq:condcomp}.
The joint running time of \textbf{Algorithm 1} and the solution of the LP problem  with 361 variables was 0.45 second on the same hardware/software environment as in the previous example. 
The obtained complex balanced and thus weakly reversible realization of the initial CRN is visible in Fig. \ref{react_02_db}. The isolated complexes (i.e. the ones with no incoming and no outgoing directed edges) are naturally omitted from the realization. It can be checked that the deficiency of the complex balanced realization is 0 (in sharp contrast with the deficiency of 12 of the initial network in Fig. \ref{react_02}). Therefore, the autonomous system \eqref{example2} has well-characterizable equilibrium points in the positive orthant that are globally stable with a known entropy-like Lyapunov function \cite{Feinberg1987}. 

An attempt to find a detailed balanced realization using the constraints described in section \ref{sec:detbal} is unsuccessful in this case. After checking the feasibility condition given in Theorem \ref{th:lpfeasibility}, we can deduce that such detailed balanced realization does not exist with the complex set listed in eq. \eqref{kinsys_complexes}.

\section{Conclusions}\label{sec:conclusions}
Polynomial-time methods were proposed in this paper for the computation of complex balanced and detailed balanced realizations of kinetic systems. The input of the methods is a given kinetic polynomial system (or an initial CRN) together with an arbitrary steady state point. The algorithms are based on standard linear programming where the characteristics of mass action dynamics are represented in the form of linear constraints. If solvable, the methods provide sufficient conditions for the existence of reversible or weakly reversible realizations in the cases of detailed balance or complex balance, respectively. The existence of such realizations usually guarantee strong stability properties for the underlying ODEs \cite{Craciun2009}. If a kinetic polynomial system is given, then a key step of the proposed methods is the algorithm described in \cite{Hars1981} for generating an initial realization. The methods can be easily combined with the computation of dense and sparse realizations \cite{Szederkenyi2011,Szederkenyi2009b}. The efficiency of the methods was demonstrated on numerical examples.
\section*{Ackowledgements}
This research work has been partially supported by the Hungarian Scientific Research Fund
 through grant no. K67625 and by the Control Engineering Research Group of the Budapest
 University of Technology and Economics.

\end{document}